\documentclass[hyper]{JHEP3}
\usepackage{epsfig}
\usepackage{amssymb}
\usepackage{amsfonts}
\usepackage{amsbsy}
\usepackage{amsmath,amssymb,amsthm,amscd,epsfig}

\def\im{\mbox{Im }}

\def\IZ{{\mathbb{Z}}}
\def\IR{{\mathbb{R}}}
\def\IP{\mathbb{P}}

\def\CM {{\cal M}}

\def\CN {{\cal N}}

\def\CO {{\cal O}}

\def\CZ {{\cal Z}}

\def\CH {{\cal H}}

\def\CZ{{\cal Z}}
\def\la{\langle}
\def\ra{\rangle}
\def\half{\frac{1}{2}}

\renewcommand{\Im}{{\rm Im }}

\def\one{{\hbox{ 1\kern-.8mm l}}}

\def\be{\bar{e}}

\def\be{\begin{equation}}
\def\ee{\end{equation}}

\def\to{\longrightarrow}

\title{Ample D4-D2-D0 Decay}

\author{Evgeny Andriyash  and Gregory~W.~Moore\\

\\
 NHETC and Department of Physics and Astronomy,
Rutgers University,\\
Piscataway, NJ 08855--0849, USA\\
\\
{\tt  andriyas@physics.rutgers.edu, gmoore@physics.rutgers.edu} }

\abstract{ We study the wall-crossing behavior  of the index of BPS
states for $D4-D2-D0$ brane systems on a Calabi-Yau 3-fold at large
radius   and point out that not only is the ``BPS index at large
radius'' chamber-dependent,  but that the changes in the index can
be large in the sense that they dominate single-centered black hole
entropy. We discuss implications for the weak coupling OSV
conjecture. We also analyze the near horizon limit of multicentered
solutions, introduced in arXiv:0802.2257, for these particular
configurations and  comment on a general criterion, conjectured in
arXiv:0802.2257, which identifies those multicentered solutions
whose near horizon limit corresponds to a geometry with  a single
asymptotic $AdS_3 \times S^2$ boundary. }

\begin{document}

\section{Introduction and Conclusion}

Consider type IIA string theory compactified on a Calabi-Yau
manifold $X$. The space of BPS states associated to D-branes
wrapping internal cycles has been a subject of much interest for
over 10 years and continues to yield surprises. In particular, for
both mathematical and physical reasons the index of BPS states of
charge $\Gamma$ has been the focus of much recent research. This
index, the second helicity supertrace of the space of BPS states,
will be denoted $\Omega(\Gamma;t)$ where  $t=B+iJ$ denotes a
complexified K\"ahler class.\footnote{We follow the notation and
conventions of \cite{Denef:2007vg}, to which we refer for further
background and references. }  Physically, we are interested in
$\Omega(\Gamma;t)$ because of its role in the Strominger-Vafa
program of accounting for black hole entropy in terms of D-brane
microstates \cite{Strominger:1996sh}. Mathematically, we expect that
$\Omega(\Gamma;t)$ will eventually be identified
 as something like ``the Euler character of the moduli
space of stable objects in the bounded derived category on $X$ with
stability condition $t$.''

At present, there are very few direct computations of the BPS
indices, and those which have been carried out are only valid in the
``large radius regime.'' That is, they assume that the K\"ahler
classes of all effective curves are large. In such a regime one can
use geometric models such as D-brane gauge theories and/or M-brane
worldvolume theories.  One can then  reduce the computation of BPS
indices to a computation in a suitable conformal field theory. This
has proven to be quite successful in a number of examples
\cite{Maldacena:1997de,Strominger:1996sh}. Therefore, let us focus
on the large radius limit of the BPS indices. To define it we choose
some vector $B+iJ$ in the complexified K\"ahler cone and consider
the limit
\begin{equation}\label{eq:largeradius}
\lim_{\Lambda\to \infty} \Omega(\Gamma; \Lambda (B+iJ) )
\end{equation}
There are of course other ways of ``going to infinity'' (for
example, a different kind of limit is taken in
\cite{AganagicJafferisMoore}) but we restrict attention to
(\ref{eq:largeradius}) in this paper. We expect - on physical
grounds - that this limit exists: In the large radius limit the
physics is described by some D-brane gauge theory, and there should
be a well-defined and finite-dimensional space of BPS states
$\CH(\Gamma;t)$. Somewhat surprisingly, it was pointed out in
\cite{Diaconescu:2007bf} that the limit (\ref{eq:largeradius})
depends on the direction $B+iJ$ chosen in the K\"ahler cone, even
for the D4-D2-D0 system studied in \cite{Maldacena:1997de}, and
hence the ``large-radius limit'' of the index of BPS states is not
well-defined without specifying more data.  This fact has recently
played an important role in \cite{deBoer:2008fk}. Our point in the
present paper is that in fact the dependence of the index on the
direction $B+iJ$ can be large and this has significant implications,
as explained in more detail below.

It turns out that $\Omega(\Gamma;t)$ is only piecewise constant as a
function of $t$, and it can jump discontinuously across walls of
marginal stability. While $\Omega(\Gamma;t)$ is difficult to compute
it turns out that there are fairly simple formulae for the
\emph{change} of $\Omega(\Gamma;t)$ across walls of marginal
stability.   The only wall-crossing formula we will need in this
paper is  the primitive wall-crossing formula of \cite{Denef:2007vg}
which states the following:  Suppose   $\Gamma_1, \Gamma_2$ are
primitive charge vectors such that $\Gamma=\Gamma_1+\Gamma_2$ then
the quantity
\begin{equation}\label{eq:denefstab}
\langle \Gamma_1, \Gamma_2 \rangle {\rm Im}(Z(\Gamma_1;t) (
Z(\Gamma_2;t))^*
\end{equation}
is positive in the stable region and negative in the unstable
region. (``Stable'' and ``unstable'' refer to the physical stability
of the multi-centered solutions of \cite{Denef:2000nb,Denef:2002ru}.
We refer to this as ``Denef stability.'' ) If we cross the marginal
stability wall, where the phases of $Z(\Gamma_1,t)$ and
$Z(\Gamma_2,t)$ are equal, from the positive side to the negative
side across a generic point $t_{ms}$ then
\begin{equation}
\Delta \Omega =(-1)^{\langle \Gamma_1, \Gamma_2 \rangle-1}\vert
\langle \Gamma_1, \Gamma_2 \rangle\vert \Omega(\Gamma_1;t_{ms})
\Omega(\Gamma_2;t_{ms}).
\end{equation}

  In \cite{Denef:2007vg} it was
pointed out that for D6-D4-D2-D0 systems there is nontrivial
wall-crossing at infinite radius. In
\cite{Diaconescu:2007bf,deBoer:2008fk} it was shown that even for
the D4-D2-D0 system with ample $D4$ charge $P$, there are walls of
marginal stability going to infinity. (Such examples are only
possible when the dimension of the K\"ahler cone is greater than one
\cite{Denef:2007vg}.) One should therefore ask how large the
discontinuities in $\Omega$ can be across walls at infinity. In this
note we show that they can be large in the following sense: If we
consider charges $\Gamma$ which support regular attractor points
(hence the single-centered attractor solutions of
\cite{Ferrara:1995ih,Strominger:1996kf}) then it is not consistent
with wall-crossing to assume that the contribution of such states
dominate the large radius limit of $\Omega$.  We show this by
exhibiting an explicit example.

Our example consists of a charge which supports a regular attractor
point (hence a single-centered black hole), but which also supports
a 3-centered solution. The three-centered solution decays across a
wall in the K\"ahler cone which extends to arbitrarily large radius.
 The
contribution of the single centered solution of charge $\Gamma$ is
predicted from supergravity to be
\begin{equation}
\log \vert \Omega \vert \sim S_{BH}(\Gamma):=2\pi \sqrt{- \frac{\hat
q_0}{6} P^3}
\end{equation}
In our example $\Gamma$ will support a boundstate of charge
$\Gamma_1+ \Gamma_2$ where $\Omega(\Gamma_1)$ has bounded entropy
and $\Gamma_2$ itself supports a regular attractor point, but
$S_{BH}(\Gamma_2) > S_{BH}(\Gamma)$. Thus the discontinuities in the index
are competitive with the single-centered entropy.

This effect of entropy dominance of multi centered configurations
over single-centered ones is similar to the ``entropy enigma''
configurations of \cite{Denef:2007vg,Denef:2007yn}. In that case, if
we first take large $J_\infty$ then under charge rescaling $\Gamma
\to \Lambda \Gamma$ single centered entropy scales as $S_{BH} \sim
\Lambda^2$ while the two-centered solutions contribute to entropy as
$S_{2c} \sim \Lambda^3$. On the other hand, if one holds the moduli
at infinity, $J_{\infty}$, fixed and scales $\Gamma$, then the
configuration will eventually become unstable and leave the
spectrum. In the example of the present note, we again first take
large $J_\infty$. Then we find that under rescaling D4 charge $P \to
\Lambda P$ (holding the remaining components of $\Gamma$ fixed) the
single centered entropy scales as $S_{BH} \sim c_{BH} \Lambda^{3/2}$
while the three-centered entropy scales as $S_{3c} \sim
c_{3c}\Lambda^{3/2}$, with $c_{3c} > c_{BH}$. Thus here the entropy
dominance of multicentered configuration arises from the prefactor
and not from the scaling exponent. In contrast to the entropy enigma
configuration, if we fix moduli at infinity $J_{\infty}$ and then
scale $P$, the configuration does not leave the spectrum, as shown
at the end of section 3 below.

Like the ``entropy enigma'' configurations, the boundstates of the
present note threaten to invalidate the weak-coupling version of the
OSV conjecture \cite{Ooguri:2004zv} (or its refined version
\cite{Denef:2007vg}). However, as discussed at length in
\cite{Denef:2007vg}, (see especially section 7.4.2), since $\Omega$
is an index there are potential cancellations between these
configurations leading to the desired scaling $\log \Omega \sim
\Lambda^2$ for uniformly scaled charges.   The point of the present
note is {\it even if we assume} that there are such miraculous
cancellations  the index will nevertheless have large
discontinuities across the MS walls, even at large radius, and hence
the weak coupling OSV conjecture is at best valid in special
chambers of the K\"ahler cone.
%
%
It is notable that the phenomenon we discuss cannot happen when the
K\"ahler cone is one-dimensional. Moreover, our example only exists
in the regime of weak topological string coupling, where $\vert \hat
q_0\vert $ is not much larger than $P^3$. This regime is already
known to be problematical for the OSV conjecture
\cite{Denef:2007vg}.

Of course, given a charge $\Gamma = P + Q+ q_0 dV$, with $P$ in the
K\"ahler cone, there is a natural direction singled out, namely the
$P$ direction. It is therefore natural to suppose that the refined
OSV formula of \cite{Denef:2007vg} should apply to
\begin{equation}
\lim_{\Lambda \to \infty}\Omega(\Gamma; \Lambda z P )
\end{equation}
where $z=x+iy$ is a complex number,  and indeed, several  of the
arguments in \cite{Denef:2007vg} assumed (for simplicity) that $J$
and $P$ are proportional.

A second, related, implication of our example concerns the
modularity of generating functions for BPS indices.  In
\cite{Denef:2007vg} a microscopic formulation of the ``large
radius'' BPS indices  was investigated by characterizing the BPS
states as coherent sheaves supported on cycles in the linear system
$\vert P\vert$. Put differently, a D4 brane wraps a cycle $\Sigma\in
\vert P\vert$. There is a prescribed flux $F\in H^2(\Sigma;\IZ)$ and
the system is bound to $N$ anti-D0 branes. If we set $d(F,N) =
(-1)^{\dim \CM} \chi(\CM)$ where $\CM$ is the moduli space of
supersymmetric configurations of this type then, it was claimed, the
large radius BPS indices are finite sums of the $d(F,N)$. On the
other hand, duality symmetries of string theory imply that   a
certain generating function of the indices $d(F,N)$, denoted
$Z_{D4D2D0}$,
 exhibits good modular behavior.   It follows
from the chamber dependence of the large radius limit of $\Omega$
that there must be chamber dependence of the $d(F,N)$.
%
%
 The chamber dependence of $d(F,N)$ raises
the question of compatibility with the modularity  of the partition
function $Z_{D4D2D0}$.  This partition function is also closely
related   to the $(0,4)$ elliptic genus of the MSW string
\cite{Gaiotto:2006ns,deBoer:2006vg}, and hence similar remarks might
apply to that elliptic genus. The statement of modularity of these
partition functions follows from very basic duality symmetries in
string theory and conformal field theory which, one might guess,
should be valid in every chamber of the K\"ahler cone. One might
therefore expect that the \emph{change} in the partition function
must also be modular. It might be easier to verify this than it is
to verify the modularity of the full partition function. One might
approach this using the results of \cite{Manschot:2007ha}: One must
compute the change of the polar polynomial across a chamber and show
that the associated cusp form vanishes. This appears to be a
challenging computation, but one well worth doing if possible.

In section \ref{sec:MTheoryLift} we check what happens to
our boundstate configurations in the near horizon scaling limit
recently introduced in \cite{deBoer:2008fk}. This is important since
our observations regarding the entropy have the potential to lead to
a troublesome contradiction with the AdS/CFT conjecture. If our
configurations corresponded to states in the Cardy region of the
 holographic dual to an asymptotically $AdS_3 \times S^2$ geometry
then there would be such a contradiction. Fortunately, our example
turns out to be   quite similar to that discussed in
\cite{deBoer:2008fk}: The first split  $D4 \to D4 + D4$ corresponds
to two infinitely separated $(AdS_3 \times S^2)$-like geometries, so
there is no contradiction. These curious limiting geometries, and
especially their holographic dual interpretation, deserve to be
understood much better. Indeed, the existence of these $D4 \to D4
+D4$ decays suggests that in general one \emph{cannot} identify the
partition function $Z_{D4D2D0}$ of \cite{Denef:2007vg} with the M5
elliptic genus of \cite{Gaiotto:2006ns,deBoer:2006vg}! They might
nevertheless agree in certain chambers of the K\"ahler cone (e.g. at
the ``AdS point'' described in \cite{deBoer:2008fk}). Clearly, this
issue deserves to be understood better.

 Finally, as a by-product of our investigation,   in
section \ref{sec:HologDuals} we discuss the general criterion,
proposed in  \cite{deBoer:2008fk}, for $D$-brane configurations to
have single asymptotic $AdS_3 \times S^2$ geometry in the near
horizon limit. We give an argument, based on the Split Attractor
Flow Conjecture, in favor of this criterion.

\section{Some general remarks on stability at large radius}

A thorough analysis of the possible walls at infinity for the D4D2D0
system, and the existence of split states in those regions is far
beyond the scope of this modest note. We would, however, like to
make a few elementary general points.

 Let us consider a D4-D2-D0 charge $\Gamma = P+Q+q_0 dV$ splitting
into a pair of charges $\Gamma = \Gamma_1+ \Gamma_2$ with
\begin{equation}
\Gamma_i = r_i + P_i + Q_i + q_{0,i} dV
\end{equation}
Then $r_1=-r_2 = r$ and $I_{12} = \langle \Gamma_1, \Gamma_2\rangle
= P_1\cdot Q_2 - P_2 \cdot Q_1 - rq_0 $. The Denef stability
condition is governed  by the sign of $I_{12}$ times the sign of
\begin{equation}\label{eq:genz12}
\CZ_{12}:=\Im Z_{1,hol} Z_{2,hol}^*.
\end{equation}

\noindent We are interested in the existence of walls at infinity.
Let us consider walls which asymptotically contain lines in the
K\"ahler moduli space. Thus, we set $t \to  \Lambda t$ and take
$\Lambda\to \infty$. If the leading term in $\CZ_{12}$ at large
$\Lambda$ can change sign as the ``direction'' $t$ is changed, then
there will be asymptotic walls at infinity.

If $r$ is nonzero then any wall that persists at infinity is
necessarily an anti-MS wall, where the phases of $Z(\Gamma_1;t)$ and
$Z(\Gamma_2;t)$ \emph{anti}-align. There is no wall-crossing
associated with such walls and  thus we set $r=0$.

When  $r=0$ (\ref{eq:genz12}) simplifies to
\begin{equation}\label{eq:genz12p}
\begin{split}
\CZ_{12}  &=    \frac{1}{4}\Im  P_1\cdot t^2 P_2\cdot \bar t^2 \\
& - \half \Im \left( P_1 \cdot t^2 Q_2\cdot \bar t + P_2\cdot \bar t^2 Q_1\cdot t \right)\\
& + \Im\left( Q_1 \cdot t Q_2 \cdot \bar t + \half q_{0,1} P_2\cdot
\bar t^2 + \half q_{0,2} P_1\cdot t^2\right) \\
& - \Im \left( q_{0,1} Q_2 \cdot \bar t + q_{0,2} Q_1 \cdot t
\right)
\end{split}
\end{equation}
For the generic direction $t$ the leading behavior for $\Lambda\to
\infty$ will be governed by the sign of
\begin{equation}\label{eq:fourthorder}
\Im  P_1\cdot t^2 P_2\cdot \bar t^2  = (P_1\cdot B \cdot J) P_2
\cdot B^2 - (P_2\cdot B \cdot J) P_1\cdot B^2 - ( P_2\cdot J^2
P_1\cdot B \cdot J - P_1 \cdot J^2 P_2 \cdot B \cdot J)
\end{equation}
This vanishes in the one-modulus case, but is   generically nonzero
in the higher dimensional cases. Moreover, it is  odd in $B$.
Therefore, just by changing the sign of $B$ we change from a region
of Denef stability to instability, and hence there are definitely
walls at infinity.

As an example we analyze (\ref{eq:fourthorder}) for   two particular
examples of Calabi-Yau manifolds with a  2-parameter moduli space.
The first case is the elliptic fibration $\pi: X \rightarrow P^2$. A
basis of divisors is $D_1 = \alpha_f, D_2 = h$ with intersection
products given by $\alpha_f^3=9$, $\alpha_f^2 h=3$, $\alpha_f h^2=1$
and $h^3=0$. The second example is a blow-up of a hypersurface in
$\IP^{(1,1,2,2,2)}[8]$ \cite{Candelas:1993}. A basis of divisors is
$H$ and $L$ with intersection products given by $H^3 = 8$, $H^2 L =
4$, $H L^2 = 0$, $L^3= 0$.  It turns out that in the elliptic
fibration case (\ref{eq:fourthorder}) takes the form (here,
superscripts denote components w.r.t. the basis $D_1$, $D_2$ above):

\begin{equation}
16 ((B^1)^2 + (J^1)^2) (P_1^2 P_2^1 -P_1^1 P_2^2) (B^2 J^1 - B^1 J^2)
\end{equation}

\noindent and thus vanishes whenever $P_1$ becomes parallel  to
$P_2$ or $B$ becomes parallel to $J$. Assuming $P_1$ not parallel to
$P_2$ there is exactly one wall, going to infinity with $B \propto
J$. In the case of  $\IP^{(1,1,2,2,2)}[8]$ (\ref{eq:fourthorder})
looks like:

\begin{equation}
(3 B_1 B_2 + B_2^2 + 3 J_1 J_2 + J_2^2) (P_1^{(2)} P_2^{(1)} -
P_1^{(1)} P_2^{(2)}) (B_2 J_1 - B_1 J_2)
\end{equation}

\noindent Here in addition to $B \propto J$ wall there is another
wall for $3 B_1 B_2 + B_2^2 + 3 J_1 J_2 + J_2^2=0$, provided that $9
B_1^2 - 12 J_1 J_2 - 4 J_2^2>0$. It is easy to  see that on the $B
\propto J$ wall the phases of the central charges align and hence,
this is an MS and not an anti-MS wall. For the additional wall,
presented above, the same is true.

It would be interesting to investigate these stable regions more
thoroughly. The stability condition is necessary, but far from
sufficient for the existence of BPS boundstates, so one cannot
immediately conclude that there is  nontrivial wall-crossing.  For
simplicity we will henceforth take   $B=0$ in this paper. In this
case the asymptotic walls are governed by the next largest term and
the stability condition at large $\Lambda$ is governed by the sign
of
\begin{equation}  (P_2 \cdot J^2 Q_1 \cdot J - P_1
\cdot J^2 Q_2 \cdot J)
\end{equation}
Again, in the one-modulus case this expression has a definite sign
in accord with the analysis in \cite{Denef:2007vg}, however, in the
higher dimensional case it is perfectly possible for this quantity
to change sign as $J$ changes direction  in the K\"ahler cone. This
is the example we will focus on.

\section{An example}\label{example}

We now give an explicit example of a split of a D4D2D0 charge, which
supports a single centered black hole, but which admits marginal
stability  walls at infinity describing a splitting  into a pair of
D4D2D0 systems in which the change in index $\Delta \Omega$ is
larger than the single-centered entropy.

In order to have a single-centered solution we must assume $P$ is in
the K\"ahler cone and the discriminant is positive. Therefore,

\begin{equation}
\hat q_0 <0 \qquad \hat q_0 := q_0 - \half Q^2\vert_P
\end{equation}
where we recall that $Q^2\vert_P :=  (D_{ABC}P^C)^{-1} Q_A Q_B$.

In some chambers this charge can also support a multicentered
solution where the first split in the attractor flow tree is given
by

\begin{eqnarray}
&& \Gamma \rightarrow \Gamma_1 + \Gamma_2 \nonumber\\
&& \Gamma_1 =  P_1+ \frac{\chi(P_1)}{24} dV \nonumber\\
&& \Gamma_2 = P_2+ Q+ q_{0,2}  dV
\end{eqnarray}

Here, $\Gamma_1$ is a pure D4-brane  and $\Gamma_2$  is a $D4$-brane
charge  supporting a single-centered black hole: We will consider
only charge configurations so that  $\hat q_{0,2}<0$, and hence
$\Gamma_2$ has a regular attractor point.

 Using the summary of split attractor flows in the appendix, we see
 that a necessary condition for the existence of the split
 realization is that  the flow crosses a
wall of marginal stability for  $\Gamma_1$ and $\Gamma_2$, at a
positive value of the flow parameter $s$. Using notations from
Appendix \ref{AttrFlTr} the flow parameter is given by:

\begin{equation}\label{eq:Sms12}
s_{ms}^{12} =  2 \frac{  -(Q \cdot J- P \cdot B \cdot J) \:( \frac{1}{2} P_1 \cdot (J^2-B^2)  + \frac{\chi(P_1)}{24})- (\frac{1}{2} P \cdot (J^2-B^2)+Q \cdot B -q_0)
\: P_1 \cdot B \cdot J  }{ \sqrt{\frac{4}{3}
J^3} |\frac{1}{2} P \cdot (J^2-B^2)+Q \cdot B -q_0 +i Q \cdot J- i P \cdot B \cdot J| \: P_1 \cdot Q}|_{\infty}
\end{equation}

\noindent Here $|_{\infty}$ means that complexified K\"ahler moduli
$t = B+iJ$ are evaluated at spatial infinity. The vanishing locus of
$s_{ms}$ is the wall of marginal stability. This is a rather
complicated expression,  but it simplifies if the starting point is
chosen to have zero $B$-field. In that case the parameter along the
flow $s_{ms}^{12}$, for which the wall is crossed is

\begin{eqnarray}\label{StableSide}
&&s_{ms}^{12} =  2 \frac{  -Q \cdot J \:( \frac{1}{2} P_1 \cdot J^2
+ \frac{\chi(P_1)}{24}) }{ \sqrt{\frac{4}{3} J^3} |\frac{1}{2} P
\cdot J^2 -i Q\cdot J -q_0| \: P_1 \cdot Q}|_{\infty}
\end{eqnarray}
which further simplifies in the large $J$ limit to
\begin{equation}\label{StableSideII}
s_{ms}^{12} =-  2 \frac{ Q \cdot J \: P_1 \cdot J^2}{
\sqrt{\frac{4}{3} J^3} P \cdot J^2 \: P_1 \cdot Q}|_{\infty}
\end{equation}
The condition   $s_{ms}^{12}>0$ (which is equivalent to the Denef
stability condition) imposes a restriction on $Q$, because we must
have $(Q J_{\infty}) (P_1 Q)<0$ while  both $P_1$ and $J_{\infty}$
are in K\"ahler cone. There are plenty of charges that satisfy this
condition and we'll give a numerical example below.

We are not quite done constructing the split attractor flow tree
because $\Gamma_1$ is a polar charge, and must itself be realized as
a multicentered solution.

As discussed in appendix \ref{AttrFlTr}, for an attractor tree to
exist all its edges must exist and moreover all its  terminal
charges must support BPS states. The charge $\Gamma_2$ supports a
regular black hole. Meanwhile,  $\Gamma_1$ is realized as a flow,
splitting into $D6$ and $\overline{ D6}$ as in \cite{Denef:2007vg}:

\begin{eqnarray}
&&\Gamma_1 \rightarrow \Gamma_3+ \Gamma_4 \nonumber\\
&&\Gamma_3 = e^{P_1/2} \nonumber\\
&&\Gamma_4 =- e^{-P_1/2} \nonumber\\
\end{eqnarray}

So for the whole tree to exist we need

\begin{itemize}
\item $s_{ms}^{12} >0$ for the split $\Gamma \to \Gamma_1 + \Gamma_2$ to exist
\item $s_{ms}^{34} >0$ for the split $\Gamma_1 \to \Gamma_3 + \Gamma_4$ to exist
\item $s_{0}^{34} >s_{ms}^{34}$ where $s_{0}^{34}$ is the value when the
flow reaches zero of the charge $Z(\Gamma_1)$
\end{itemize}

These conditions are sufficient because the charges $\Gamma_3$ and
$\Gamma_4$ exist everywhere in moduli space and $\Gamma$ and
$\Gamma_2$ support black holes. It is also easy to see that both
walls are MS and not anti-MS walls. It turns out that above
conditions are always satisfied if

\begin{itemize}
\item $J_{\infty}$ is on stable side of the wall, corresponding to $s_{ms}^{12}>0$
\item $P_1 \ll J_{\infty}$ component-wise in a basis of K\"ahler cone
\end{itemize}


To see this we estimate $s_{ms}^{34}$ and $s_{0}^{34}$ in the large
$J_{\infty}$ limit. Recall from appendix \ref{AttrFlTr} that
\begin{equation}
s_{ms}^{34} = \frac{\la\Gamma_3, \Delta H \ra - \la \Gamma_3, \Gamma
\ra s_{ms}^{12}}{\la \Gamma_3, \Gamma_4 \ra}.
\end{equation}
Now plugging the expression for $\Delta H$ from  (\ref{FlowModuli})
we can estimate $\la \Gamma_3, \Delta H \ra \sim
\frac{J_{\infty}^3}{3 \sqrt{4/3 J_{\infty}^3}}$. Using $\la
\Gamma_3, \Gamma_4 \ra = -\frac{P_1^3}{6}$ and the fact that
$s_{ms}^{12} \sim O(\frac{1}{J^{1/2}})$ is small  we get
\begin{equation}\label{sms34J}
s_{ms}^{34} \sim \frac{2 J_{\infty}^3}{ \sqrt{4/3 J_{\infty}^3}
P_1^3 }.
\end{equation}
To find $s_{0}^{34}$ we equate the central charge to zero
$Z(\Gamma_1; t)=0$ to get the vanishing locus:
\begin{eqnarray}\label{eq:CentralCharge}
-\frac{\chi(P_1)}{24} -\frac{1}{2} P_1 \cdot B^2 + \frac{1}{2}
P_1 \cdot J^2 =0, \qquad  P_1 \cdot B \cdot J=0
\end{eqnarray}

Moduli along the flow of charge $\Gamma_1$ are determined  again by
(\ref{FlowModuli}) with $\Gamma(s) = s \Gamma_1 + s_{ms}^{12} \Gamma
- \Delta H$. Recalling that $s_{ms}^{12} \sim O(\frac{1}{J^{1/2}})$
this can be written as
\begin{equation}
\Gamma(s) =  \left(O(\frac{1}{J_{\infty}^{5/2}}) ,s P_1 +
O(\frac{1}{J_{\infty}^{1/2}}), O(\frac{1}{J_{\infty}^{1/2}}) , s
\frac{\chi(P_1)}{24} - \frac{J_{\infty}^3}{2 \sqrt{\frac{4}{3}
J_{\infty}^3}}\right)
\end{equation}
Plugging this $\Gamma(s)$  into (\ref{FlowModuli}) and  taking into
account that $s_0^{34} \sim O(J^{3/2})$, as we will see below, we
find that

\begin{equation}
J(s_0^{34})^a \sim P_1^a \sqrt{\frac{-6}{P_1^3}  \left(
\frac{\chi(P_1)}{24} - \frac{J_{\infty}^3}{2 s_0^{34}
\sqrt{\frac{4}{3} J_{\infty}^3})} \right)}
\end{equation}

\noindent and $B^a(s_0^{34})$ is small. Now we can  solve
(\ref{eq:CentralCharge}) for $s_0^{34}$ to find:

\begin{equation}\label{s034J}
s_{0}^{34}  \sim \frac{6 J_{\infty}^3}{ \sqrt{4/3
J_{\infty}^3}(P_1)^3 }
\end{equation}

\noindent Thus we see from (\ref{sms34J}) and (\ref{s034J}) that the existence
conditions are indeed satisfied: $s_{0}^{34} >s_{ms}^{34}$.

We conclude with a numerical example, checking explicitly  that such
split solutions exist.  We consider again the elliptic fibration
example and $\IP^{(1,1,2,2,2)}[8]$ of \cite{Candelas:1993}.

The initial charge is of the form $\Gamma = P+Q +q_0 dV$, where
$P=(50,50)$, $Q=(-1,3)$, $q_0 = -10$. The starting point of the flow
is $J_{\infty} = (500,100)$, which indeed lies on stable side of MS
wall in (\ref{StableSide}). The pure $D4$ has charge $P_1 = (1,2)$.
All the existence conditions are found  to be satisfied for both
Calabi-Yau manifolds. As we'll discuss in the next section, the
entropy of this three-centered configuration is expected to be
larger than the one from the single-centered realization of the same
total charge. The numerical examples confirm this claim in both
cases.

Now we will justify the remark made in the introduction about the
existence of the split state for $P \to \infty$. We take
$B_{\infty}=0$ and evaluate (\ref{eq:Sms12}). Evaluating
(\ref{StableSide}) in the limit $P \to \infty$ and with fixed
$J_{\infty}$ produces an expression almost identical to
(\ref{StableSideII}). In particular, it
 remains positive, but does go to zero. The second
split $\Gamma_1 \to \Gamma_3 + \Gamma_4$ will therefore happen very
close to starting point in moduli space and hence  $J_{\infty} \gg
P_1$  will guarantee that   the second split   exists. This proves
that our example exists in the $P \to \infty$ limit if it existed in
$J_{\infty} \to \infty$ limit.

\section{Comparison of the entropies}

Now let us compare the discontinuity $\Delta \Omega$ of the BPS
index with the contribution of the single-centered (black hole)
solutions to the ``large radius'' index $\Omega(\Gamma;J_{\infty})$.

We first \emph{assume} that the dominant contribution to the large
radius entropy is that of the single-centered solutions, if they
exist. We will then show that this assumption is inconsistent with
the wall-crossing phenomena.

The black hole contribution to $\Omega$  can be approximated  using
the equation from the attractor mechanism

\begin{equation}\label{BHentrG}
\Omega_{BH}(\Gamma) :=  \exp S_{BH}(\Gamma) =  \exp\left[2 \pi \sqrt{-\hat q_0 P^3/6}\right]
,
\end{equation}

The discontinuity of the index across the wall  $\Gamma \rightarrow
\Gamma_1+\Gamma_2$ is given by

\begin{equation}\label{Delta12}
\Delta_{12} \Omega(\Gamma; t_{ms}) = (-1)^{\la \Gamma_1,\Gamma_2
\ra -1} | \la \Gamma_1,\Gamma_2 \ra | \: \Omega(\Gamma_1; t_ {ms}^{12}) \:
\Omega(\Gamma_2; t_{ms}^{12})
\end{equation}

\noindent Here the  indices of $\Gamma_1$ and $\Gamma_2$ are
evaluated on the MS wall. As we have said, the state with charge
$\Gamma_1$ is realized as a split attractor flow splitting into pure
$D6$ and $\overline{ D6}$ with fluxes. The index of $\Gamma_1$ is
polynomial in charges and is given by $\Omega(\Gamma_1)
=(-1)^{I(P_1)-1} I(P_1)$ where $I(P) :=  \frac{P ^3}{6} +\frac{
c_2(X) \cdot P }{12}$. Again using our assumption we would estimate
that the index of $\Gamma_2$ can again be approximated by the  black
hole contribution:
\begin{equation}\label{BHentrG2}
\Omega(\Gamma_2, J_\infty) \sim \Omega_{BH}(\Gamma_2)=  \exp\left[2
\pi \sqrt{-\hat q_{0,2} P_2^3/6}\right]
\end{equation}
since $\Gamma_2$ supports a single-centered black hole.

We now  consider a limit of large charges. We hold $P_1$ fixed
and take $P\to \infty$ along some direction in the K\"ahler cone. Then
from Eqs.(\ref{BHentrG}), (\ref{BHentrG2}) the indices of $\Gamma$
and $\Gamma_2$ will be exponentially large for large $P$ while
$\Omega(\Gamma_1)$ is a known, bounded function of $P_1$.  This
means that to compare the contributions (\ref{BHentrG}) and
(\ref{Delta12}) we need to compare the exponents:

\begin{equation}
-\hat q_0 P^3  \quad {\rm vs} \quad -\hat q_{0,2} P_2^3
\end{equation}

In this  limit we can write
\begin{equation}
P_2^3 = P^3 - 3 P^2 \cdot P_1 + ... = P^3 \left(1 - \frac{3
P^2 \cdot P_1}{P^3} + \CO(1/\vert P \vert^2)\right).
\end{equation}
Moreover, since $q_0$ is conserved at the vertex
\begin{equation}
\hat q_{0,2} = \hat q_0 + \frac{1}{2}Q^2|_P  -
\frac{\chi(P_1)}{24} -\frac{1}{2} Q^2|_{P_2}
\end{equation}
In taking our charge limit we can make $q_{0,2}$ sufficiently
negative that $\hat q_{0,2}$ and $\hat q_0$ are both negative. Now
we can write
\begin{equation}\label{ExpComp}
- \hat q_{0,2} P_2^3 = -\hat q_0  P^3  \left(1  -
\frac{\chi(P_1)}{24 \hat q_0} - \frac{1}{2 \hat q_0}
(Q^2|_{P_2} - Q^2|_P) - \frac{3 P^2 \cdot P_1}{P^3} +
\CO(1/\vert P \vert^2) \right)
\end{equation}

Since $\hat q_0$ is negative we see from   (\ref{ExpComp})  that the
contribution of the  $\Gamma \to \Gamma_1+\Gamma_2\to (\Gamma_3 +
\Gamma_4) + \Gamma_2$ split attractor flow will be greater in the
$P\to \infty$ limit  provided that

\begin{equation}\label{eq:compare}
\frac{\chi(P_1)}{24\vert \hat q_0\vert } + \frac{1}{2\vert \hat
q_0 \vert } (Q^2|_{P_2} - Q^2|_P) - \frac{3 P^2 \cdot
P_1}{P^3}
>0
\end{equation}

%

The first term of (\ref{eq:compare})   is always positive, while the
second term  can have both signs. The third term is always negative.
However, for parametrically large $P$ and fixed $Q$ the second and third terms are
 suppressed, so the expression is positive.  Thus we find that in the
limit described above, the split flow configuration has {\it
greater} entropy than the  black hole contribution:

\begin{equation}\label{eq:dominance}
  \Omega_{BH}(\Gamma )  \ll  \Delta_{12}
\Omega(\Gamma; t_{ms}).
\end{equation}

Thus, as explained in the introduction, not only does the value of
the index $\Omega$ depend on the direction in which $J$ is taken to
infinity, but this dependence can be very strong, and even dominate
single-centered black hole entropy.

One might worry that there are other split flow realizations of the
charge $\Gamma$, with the same wall of marginal stability as the one
we are studying, which produce a cancellation in $\Delta \Omega$.
For example, the charge $\Gamma_2$ might well support multi-centered
solutions. However, by our hypothesis, the single-centered entropy
dominates the multi-centered ones, so such a cancellation cannot
occur. Then (\ref{eq:dominance}) leads to a contradiction and hence
we conclude that it cannot be that single-centered entropy dominates
the entropy at infinity in all chambers.

\bigskip
\textbf{Remarks}

\begin{enumerate}

\item In the context of     topological string theory the
topological string coupling $g_{top} \sim \sqrt{-\hat q_0/P^3}$
\cite{Ooguri:2004zv}. The effect we are discussing does not appear
in the strong coupling regime, in harmony with the arguments in
\cite{Denef:2007vg}. However, it does appear in the problematic weak
coupling regime.

\item Interestingly, this phenomenon will not occur with splits into
two single-centered attractors. If $q_{0,i}<0$ for both $i=1,2$
and $P_1, P_2$ are in the K\"ahler cone then (taking
$Q_i=0$ for simplicity)  one can show that
\begin{equation}
S_{BH}(\Gamma)> S_{BH}(\Gamma_1) + S_{BH}(\Gamma_2)
\end{equation}
as expected. We do not know of a proof of the analogous statement for $Q_i \ne 0$.
\item

In principle the example we have given can be extended by replacing $\Gamma_1$ by an arbitrary extreme polar state in the sense of \cite{Denef:2007vg}. Following \cite{Denef:2007vg}, the charges $\Gamma_1 \to \Gamma_3 + \Gamma_4$ can be parametrized as

\begin{eqnarray}
&& \Gamma_3 = r e^{S_1} (1-\beta_1 + n_1 w) \nonumber\\
&& \Gamma_4 = -r e^{S_2} (1-\beta_2 + n_2 w) \nonumber\\
&& \Gamma_1 = r \left( 0, \hat P, \frac{\hat P S}{2}+ \Delta \beta,
\frac{\hat P^3}{24}+ \frac{\hat P S^2}{8} -\frac{\hat P \beta}{2}
+ \frac{S \Delta \beta}{2} - \Delta n w\right)
\end{eqnarray}

\noindent where $\hat P = S_1-S_2$, $S=S_1+S_2$, $\beta =
\beta_1+\beta_2$, $\Delta \beta = \beta_2-\beta_1$, $\Delta n =
n_2-n_1$. For  sufficiently small $\beta_i$ and $n_i$ and $S_1
\equiv P_1/2$, $S_2 \equiv -P_1/2$, the charge $\Gamma_1$ is very
close to a pure $D4$-brane and all existence conditions are still
satisfied. The 3-centered entropy dominance also continues to hold.

\end{enumerate}

\section{M-theory lift and its near-horizon limit}
\label{sec:MTheoryLift}

In this section we discuss the M-theory lift  of the 3-centered
configuration of interest and analyze its near horizon limit
following the procedure of \cite{deBoer:2008fk}. Our motivation here
is to relate our configurations to the MSW conformal field theory,
and to check that there is no contradiction with AdS/CFT.

  The solution to the attractor equations in the effective 4d $\CN=2$ SUGRA for a general
multicentered configuration can be written (in the regime of large
K\"ahler classes) in terms of harmonic functions (
\cite{deBoer:2008fk}, eq. (2.8)):

\begin{eqnarray}\label{eq:AttrSol}
 ds^2_{4d} & = & -\frac{1}{\Sigma}(d x_0+\sqrt{G_4}\,\omega)^2+\Sigma \, (d\vec x)^2\,,
\nonumber\\
{\cal A}^0 & = & \frac{\partial \log \Sigma}{\partial H_0}\left(\frac{d x_0}{\sqrt{G_4}}+\omega \right)+\omega_0\,,\label{multicenter}\nonumber\\
{ \cal A}^A & = & \frac{\partial \log \Sigma}{\partial H_A}\left(\frac{d x_0}{\sqrt{G_4}}+\omega \right)+{ \cal A}_d^A\,,\nonumber\\
 t^A&=&\frac{H^A}{H^0}+\frac{y^A}{Q^{\frac{3}{2}}}\left(i\Sigma-\frac{L}{H^0}\right),
\end{eqnarray}

\noindent where

\begin{eqnarray}
 \star d\omega & = & \frac{1}{\sqrt{G_4}} \langle dH, H\rangle \, , \quad d\omega_0 =  \frac{1}{\sqrt{G_4}}\star dH^0 \,
\nonumber\\
d{\cal A}_d^A & = & \frac{1}{\sqrt{G_4}}\star dH^A \, , \quad \Sigma = \sqrt{\frac{Q^3-L^2}{(H^0)^2}}\,
\nonumber\\
 L & = & H_0(H^0)^2+\frac{1}{3}D_{ABC}H^AH^BH^C-H^AH_AH^0\,, \label{conditions}\\
 Q^3 & = & (\frac{1}{3}D_{ABC}y^Ay^By^C)^2\,, \quad  D_{ABC}y^Ay^B  =  -2H_CH^0+D_{ABC}H^AH^B\,
\nonumber\\
H &\equiv & (H^0,H^A,H_A,H_0) :=  \sum_a \frac{\Gamma_a \,
\sqrt{G_4}}{|\vec{x}-\vec{x}_a|}-2 \im(e^{-i\alpha}\Omega)|_{\vec
x=\infty} \, ,   \nonumber
\end{eqnarray}

\noindent   $A=1,\dots, h^{1,1}(X)$ are components relative to a
basis $D_A$ for $H^2(X,\IZ)$, $\star$ is the Hodge star with respect
to the  Euclidean metric $d \vec x^2$ on $\IR^3$, and we choose a
solution $y^A$ of the quadratic equations such that $y^A D_A $ is in
the K\"ahler cone. The
  Calabi-Yau volume in   string units is given by
  \begin{equation}
\tilde{V}_{IIA} =\frac{D_{ABC}}{6}J^AJ^BJ^C=\frac{1}{2}
\frac{\Sigma^3}{Q^3}
\end{equation}
and $G_4$ is the $4$-dimensional Plank constant, determined in terms
of the string length $l_s$ and string coupling $g_s$ by
\begin{eqnarray}\label{eq:G_4Vs}
G_4 = \frac{l_s^2 g_s^2}{32 \pi^2 \tilde V_{IIA,\infty}}.
\end{eqnarray}
The above equations assume $H^0(\vec x)$ is nonzero, but they have a
smooth limit as $H^0 \to 0$. (See \cite{Moore:1998pn} eq. (9.21) for
the relevant expansions.)

This solution of 4d supergravity can be lifted to 5d supergravity. To do this we use the standard relation between $M$-theory and IIA geometries
\begin{eqnarray}\label{eq:5dsol}
ds^2_{5d}&=&\frac{R^2}{4} e^{\frac{4}{3} \phi}
\left(d\psi+{\cal A}^0\right)^2+e^{-\frac{2}{3} \phi}\,ds^2_{4d}\,,\nonumber\\
Y^A&=&\tilde{V}_{IIA}^{-1/3}\,J^A\, , \quad A_{5d}^A={\cal A}^A+B^A\left(d\psi+{\cal A}^0\right)\,.
\end{eqnarray}
 Here $R$ is the M-theory circle radius,  $\psi \sim \psi + 4\pi $, $Y^A$
are 5d SUGRA moduli, and $\phi(\vec x)$ is the 10d dilaton field,
normalized as $\phi(\infty)=0$. Note that the Calabi-Yau volume in
11d Planck units is
\begin{equation}\label{eq:Mvolume}
 \tilde V_{M} = e^{-2 \phi} \frac{\tilde V_{IIA}}{g_s^2}.
\end{equation}

The near horizon limit of the $M$-theory solution, introduced in
\cite{deBoer:2008fk}, may be described as follows. Beginning with a
solution (\ref{eq:AttrSol}) we introduce a  family of BPS solutions
of the 4d supergravity equations, parametrized by $\lambda \in
[1,\infty)$. The expressions that get modified under this
deformation  are given by
\begin{eqnarray}\label{eq:family1}
ds^2_{4d \, , \lambda} & = & -\frac{1}{\Sigma^{\lambda}}(d x_0+\lambda^{-3/2} \sqrt{G_4}\,\omega^{\lambda})^2+ \lambda^{-6}\Sigma^{\lambda} \, (d\vec x)^2\,,
\nonumber\\
{\cal A}^0_{\lambda} & = & \frac{\partial \log \Sigma^{\lambda}}{\partial H_0^{\lambda}}\left( \lambda^{3/2} \frac{d x_0}{\sqrt{G_4}}+\omega^{\lambda} \right)+\omega_0^{\lambda}\,,
\nonumber\\
\star d\omega^{\lambda} & = & \frac{\lambda^{-3/2}}{\sqrt{G_4}} \langle dH^{\lambda}, H^{\lambda}\rangle \, , \quad d\omega_0^{\lambda} =  \frac{\lambda^{-3/2}}{\sqrt{G_4}}\star dH^0_{\lambda} \,
\nonumber\\
H^{\lambda}& :=& \lambda^{3/2}\sum_a \frac{\Gamma_a \,
\sqrt{G_4}}{|{\vec x}-{\vec x}_a^\lambda|} -2 \Im (e^{- i \alpha} \Omega)|_{B_{\infty}+i \lambda J_{\infty}}
\end{eqnarray}
Here, $\Omega= -\frac{1}{\sqrt{4/3 J^3}} e^{B+iJ}$ and for brevity
we omit the corresponding formulae for ${\cal A}^A_{\lambda}$ and
${\cal A}^A_{d \, , \lambda}$. The vectors $\vec x_a^\lambda$ used
to define $H^\lambda$ can be taken to be any   solution of the
integrability constraints
\begin{equation}\label{eq:integrability}
\,\sum_{b \neq a}  \frac{\langle \Gamma_a,\Gamma_b
\rangle}{x_{ab}^{\lambda} }= - \lambda^{-3}\sqrt{\frac{3}{G_4
J_\infty^3}}\im\left(e^{-i\alpha_{\infty,\lambda}} \int \Gamma_b
e^{-(B_\infty + i \lambda J_\infty)}\right)\qquad \forall b.
\end{equation}
where $x_{ab}^\lambda:= \vert \vec x_a^\lambda - \vec
x_b^\lambda\vert$ and $e^{i \alpha_{\infty,\lambda}}$ is the phase
of the total central charge at $B_\infty + i \lambda J_{\infty}$. We
choose $\vec x_a^\lambda$ to coincide with our original solution at
$\lambda =1$, and let them depend continuously on $\lambda$. Clearly
there is some degree of arbitrariness at this stage. \footnote{In
principle some components of the moduli space of solutions to
(\ref{eq:integrability}) might be obstructed by the positivity of
the discriminant. }

The above family of solutions can be obtained from original ones by scaling (\ref{conditions})
\begin{eqnarray}\label{eq:scaling}
&& \vec x \to \lambda^{-3} \vec x \nonumber\\
&& \l_s \to \lambda^{-3/2} \l_s \nonumber\\
&& g_s^2 \to  \lambda^{3} g_s^2 \nonumber\\
&& G_4 \to  \lambda^{-3} G_4 \nonumber\\
&& J_{\infty} \to \lambda  J_{\infty}\nonumber\\
&& B_{\infty} \to B_{\infty}
\end{eqnarray}
but we prefer to keep $\vec x, l_s, G_4 $ fixed and change the
solution according to (\ref{eq:family1}). The constant $G_4$, and
the coordinate system,  in these equations is $\lambda$-independent.

Now consider the corresponding $\lambda$-deformed 5d geometries.
Since the moduli $t^A(\vec x; \lambda)$ determined by
(\ref{eq:AttrSol})   scale as $\lambda^0$ for $\lambda \to \infty$
(at least when $H^0(\vec x)\not=0$)  it is clear that if the $\vec
x_a^\lambda$ have a well-defined limit then there are well-defined
limiting moduli $\tau^A(\vec x) := \lim_{\lambda\to \infty} t^A(\vec
x;\lambda)$. One must be careful because the limits $\vec x \to
\infty $ and $\lambda \to \infty$ do not commute. Indeed $t^A(\vec
x; \lambda) \to B^A_\infty+i \lambda J^A_\infty$ as $\vec x \to
\infty $ for any fixed $\lambda$ while $\tau^A(\vec x)$
 has  asymptotics for large $x=\vert \vec x \vert$:
\begin{eqnarray}\label{eq:asympmoduli}
\tau^A =  D^{AB} Q_B + \CO(1/x) +    i \sqrt{\frac{3 |x|}{P^3}}
(J_{\infty}^3/3)^{1/4} P^A \left(1 + \CO(1/x) \right)
\end{eqnarray}
This implies that the 5d SUGRA moduli $Y^A(\vec x)$ have
well-behaved large $\vec x$ asymptotics
\begin{equation}
Y^A(\vec x) = \frac{P^A}{(P^3/6)^{1/3}} + \CO(1/\vert \vec x \vert).
\end{equation}
Moreover,  since the 10d dilaton scales according to
(\ref{eq:Mvolume}) as $ e^{2 \phi^{\lambda}}  = \frac{\tilde
V_{IIA}^{\lambda}}{\lambda^3 g_s^2 \tilde V_{M}}$ ($\tilde V_{M}$ is
$\lambda$ independent), $e^{2 \phi^{\lambda}(\vec x)}$  for fixed
$\vec x$ scales as $\lambda^{-3}$. Note, however, that in the other
order of limits $\phi^{\lambda}(\infty)=0$. The corresponding 5d
metric for the deformed solution $\lambda^{2} ds^2_{5d ,\, \lambda}$
has a well-defined limit. Reference
 \cite{deBoer:2008fk} shows that this limiting   solution
  defines a geometry which is asymptotically
$AdS_3 \times S^2$, where there is a nontrivial connection
on the (trivial) $S^2$ bundle over the asymptotic $AdS_3$ region.

The upshot is that \emph{if} we can choose the centers $\vec
x_a^\lambda $, constrained by (\ref{eq:integrability}), so that the
$\vec x_a^\lambda$ have a well-defined finite limit as $\lambda \to
\infty$ then, by AdS/CFT, the BPS states corresponding to the
multicentered solution at $\lambda=1$ should correspond to BPS
states in the MSW conformal field theory. However, it can happen
that as $\lambda \to \infty$ the distances between the centers $\vec
x_a^\lambda$ cannot remain bounded. In this case the behavior of the
limiting geometry is more complicated, and might involve, for
example, ``several $AdS_3 \times S^2$ geometries at infinite
separation.'' In particular, note that if the total $D6$  charge
vanishes then $\alpha_{\infty,\lambda} \to 0$ and hence those
integrability equations (\ref{eq:integrability}) with
$\Gamma_{b}^0=0$  have a zero on the RHS. This might force some
centers to move to infinity.

In view of the above results we next  turn to our 3-centered
configuration and examine the integrability conditions on the
positions of the three   centers. For the set of charges  described
in section 3  we have two independent equations:

\begin{eqnarray}\label{eq:intexample}
&&  -\frac{\langle  \Gamma_2,\Gamma_3 \rangle}{x_{23}^{\lambda}}
+  \frac{\langle  \Gamma_3,\Gamma_4 \rangle}{x_{34}^{\lambda}}  = \theta_3^{\lambda}\nonumber\\
&&  -\frac{\langle  \Gamma_2,\Gamma_4 \rangle}{x_{24}^{\lambda}}  -
\frac{\langle  \Gamma_3,\Gamma_4 \rangle}{x_{34}^{\lambda}}  =
\theta_4^{\lambda}
\end{eqnarray}
where  $\theta_b^{\lambda}$ denote (minus) the right-hand-sides of
(\ref{eq:integrability}).
%
The intersections of charges take the form:

\begin{eqnarray}
&& \langle  \Gamma_3,\Gamma_4 \rangle
=- \frac{P_1^3}{6} := c \nonumber\\
&& \langle  \Gamma_2,\Gamma_3 \rangle =\left( \frac{P \cdot
P_1^2}{8} -  \frac{P_1^3}{8} + q_{0,2}  \right)
- \frac{Q \cdot P_1}{2} := a-b\nonumber\\
&& \langle  \Gamma_2,\Gamma_4 \rangle = -\left( \frac{P \cdot
P_1^2}{8} -  \frac{P_1^3}{8} + q_{0,2} \right) - \frac{Q \cdot
P_1}{2} := -a-b
\end{eqnarray}
Using the charges of section 3 and the limit $P \to \infty$  holding
$P_1$ fixed, we have $a \gg b, c$ and  $c<0$. As for the sign of $b$
we first choose $b>0$ and explain the case $b<0$ later. Equations
(\ref{eq:intexample}) determine $  x_{23}^{\lambda} $ and $
x_{24}^{\lambda} $ in terms of $  x_{34}^{\lambda}  $. As discussed
above, there is still freedom in choosing the dependence of $
x_{34}^{\lambda}  $ on $\lambda$. One way to fix this freedom is to
choose $  x_{34}^{\lambda}  $ independent of $\lambda$. The
relations between $ x_{ab}^\lambda$, following from
(\ref{eq:intexample}) are subject to the triangle inequalities. The
moduli space of solutions will  generically consist of several
intervals on the $  x_{34}^{\lambda}  $ line. The relation between
these intervals and topologies of attractor flow trees is the
essence of the Split Attractor Flow Conjecture (SAFC)
\cite{Denef:2000nb}, which we recall in  Appendix A for convenience.

\begin{figure}[htp]
\centering
\includegraphics[height=4cm,angle=0,trim=0 0 0 0]{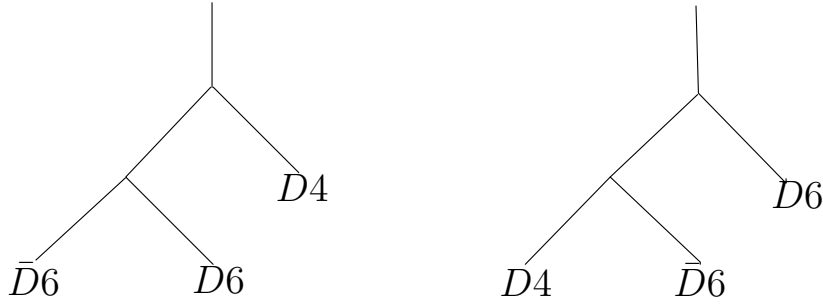}
\caption{The two contributing topologies of attractor trees. The
left tree is the main example of this paper. The right tree also
exists, for our charges, in certain regions of moduli space. }
\end{figure}


In the present case the two possible attractor flow tree topologies
are   shown in Figure 1. To identify the region corresponding to the
left tree, we tune the moduli at infinity to be close to the $D4 \to
D4 + D4$ MS wall. This means choosing $\theta_2^{\lambda} =
-(\theta_3^{\lambda}+\theta_4^{\lambda})$ close to zero. We can then
write the triangle inequalities as follows:

\begin{eqnarray}\label{eq:ineq}
&& \frac{a-b}{c-\theta_3^{\lambda} x_{34}} + \frac{a+b}{c+\theta_4^{\lambda} x_{34}} \ge 1\nonumber\\
&& \frac{a-b}{c-\theta_3^{\lambda} x_{34}} +1 \ge  \frac{a+b}{c+\theta_4^{\lambda} x_{34}} \nonumber\\
&& 1 + \frac{a+b}{c+\theta_4^{\lambda} x_{34}} \ge \frac{a-b}{c-\theta_3^{\lambda} x_{34}}
\end{eqnarray}

\noindent Close to the MS wall $\theta_2^{\lambda}=0$,  we can write
$\theta_3^{\lambda} = -\theta_4^{\lambda}-\theta_2^{\lambda}$, solve
inequalities (\ref{eq:ineq}) and expand the solution to first order
in $\theta_2^{\lambda}$. Using in addition the relations between the
magnitudes of $a,b,c$, we get the following solutions to
(\ref{eq:ineq}):

\begin{eqnarray}
&&  -\frac{c}{\theta_4^{\lambda}} + \frac{c }{2
(\theta_4^{\lambda})^2} \theta_2^{\lambda} \le x_{34} \le \frac{2
a}{\theta_4^{\lambda}}- \frac{c}{\theta_4^\lambda}
- \frac{a }{(\theta_4^{\lambda})^2} \theta_2^{\lambda} \nonumber\\
&& x_{34} \le \frac{-2 b -c}{\theta_4^{\lambda}} - \frac{a (2b +c) }{2 b(\theta_4^{\lambda})^2} \theta_2^{\lambda}  \quad {\rm or} \quad  -\frac{c}{\theta_4^{\lambda}} + \frac{a c }{2 b (\theta_4^{\lambda})^2} \theta_2^{\lambda} \le x_{34} \nonumber\\
&&  x_{34} \le -\frac{c}{\theta_4^{\lambda}} + \frac{a c }{2 b (\theta_4^{\lambda})^2} \theta_2^{\lambda}  \quad {\rm or} \quad  \frac{2 b -c}{\theta_4^{\lambda}} + \frac{a (2b-c) }{2 b (\theta_4^{\lambda})^2} \theta_2^{\lambda} \le x_{34} \nonumber\\
\end{eqnarray}

\noindent It is easy to see from these inequalities  that for
$\theta_2^{\lambda}<0$ the solution consists of a point and an
interval:

\begin{equation}
x_{34} \in \{ -\frac{c}{\theta_4^{\lambda}}  + \frac{a c }{2 b
(\theta_4^{\lambda})^2} \theta_2^{\lambda}   \} \bigcup \{ \frac{2 b
-c}{\theta_4^{\lambda}} + \frac{a (2 b-c)}{2 b (\theta_4^{\lambda})^2}
\theta_2^{\lambda}, \:  \frac{2 a}{\theta_4^{\lambda}} - \frac{a
}{(\theta_4^{\lambda})^2} \theta_2^{\lambda} \}.
\end{equation}

\begin{figure}[htp]
\centering
\includegraphics[height=2cm,angle=0,trim=0 0 0 0]{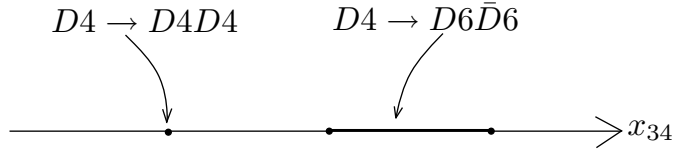}
\caption{The two intervals, corresponding to topologies of Figure 1.}
\end{figure}

\noindent On the other hand for $\theta_2^{\lambda}>0$ the point
disappears, and the solution is just an interval.
 Thus, under the SAFC correspondence, the attractor tree topology of
 our main example is identified with the component
 of the moduli of solutions to (\ref{eq:intexample}),
 given by the point on the $x_{34}^{\lambda}$ line.
 In the above we have chosen a definite sign of $b$, but it is easy to
 check that choosing $b<0$ would lead to the existence of a point
 for $\theta_2^{\lambda}>0$, and absence of it for $\theta_2^{\lambda}<0$.
 This can also be seen from the stability condition for the $D4\to D4D4$ split,
  $ -\frac{\theta_2}{\langle  \Gamma_1,\Gamma_2 \rangle}>0$, taking
  into account $\langle  \Gamma_1,\Gamma_2 \rangle = 2 b$.

Having identified the  intervals with the corresponding topologies
we can investigate   what happens to each interval as we change
$\lambda$ from $1$ to $\infty$. From the functional form of
$\theta_a^{\lambda}$ it is easy to see that $\theta_2^{\lambda}
=\CO(\lambda^{-2})$ and $\theta_{4}^{\lambda} =O(1) $ as $\lambda
\to \infty$. Thus in the near horizon limit the point on the  $\vert
\vec x_{34}^{\lambda}\vert $ line corresponding to the topology of
interest  goes to $\vert \vec x_{34}^{\lambda}\vert  =
-\frac{c}{\theta_4^{\infty}}$. This means that $\vec
x_{23}^{\lambda}, \vec x_{24}^{\lambda} \to \infty$ as $\lambda \to
\infty$ and we get an infinite separation between charges $\Gamma_2$
and $\Gamma_3+\Gamma_4$.

The conclusion is that our 3-centered configuration  {\it does not}
correspond to a single smooth geometry with $AdS_3 \times S^2$
asymptotics  in the near horizon limit of \cite{deBoer:2008fk}. This
is just as well, as pointed out in the introduction.

\section{Some general remarks on holographic duals of $D4D4$ boundstates.}\label{sec:HologDuals}

As a byproduct of our investigation of the previous section we would
like to make some more general remarks concerning the relation
between the split attractor flows and the existence of a near
horizon   geometry with a single $AdS_3 \times S^2$ boundary. In
\cite{deBoer:2008fk} it is stated that configurations with the first
split of the type $D4 \to D4+D4$ do not correspond to geometries
with a  single $AdS_3 \times S^2$ boundary. In this section we will
refine this statement.  We begin with the integrability conditions:

\begin{equation}\label{eq:integrability1}
\,\sum_{b \neq a}  \frac{\langle \Gamma_a,\Gamma_b
\rangle}{x_{ab} }= \theta_a \qquad \theta_a:= 2 \Im (e^{-i \alpha} Z(\Gamma_a))_{\infty}
\end{equation}

\noindent and denote by $M(\theta)$   the moduli space of solutions
in $\vec x_a$  to (\ref{eq:integrability1}). The decomposition of
the charges in the first split defines a disjoint decomposition of
the charges into two sets  $A\amalg B$.   Then, summing
(\ref{eq:integrability1}) over all charges in one cluster  we get:

\begin{equation}\label{eq:summedint}
\,\sum_{a \in A, b \in B}  \frac{\langle \Gamma_a,\Gamma_b
\rangle}{x_{ab} }= \theta_A :=   2 \Im (e^{-i \alpha}
Z(\Gamma_A))_{\infty}
\end{equation}

{\em \noindent \textit{ {\bf Conjecture 1:} The component of
$M(\theta)$ that corresponds to a topology with the first split $D4
\to D4+D4$ according to $A\amalg B$ under the SAFC, has the
property:
if $\sum_{a \in A} \theta_a \to 0$, then $x_{ab} \to \infty$ for $\forall a \in A \, , b \in B$.
}}

We do not know the proof of this statement  but our previous
3-centered example can serve as an illustration of it. A suggestive
argument here is  the following: Tune the moduli at infinity
$t_{\infty}$ close to the MS wall of the first split. Then,
according to the SAFC, for the $D4 \to D4D4$ component of moduli
space the $D4$ clusters will become separated, and denoting the
maximum size of these clusters by $d$, we can write
(\ref{eq:summedint}) as
\begin{equation}
 \frac{\langle \Gamma_A,\Gamma_B
\rangle}{r_{AB} } \left(1+O(\frac{d}{r_{AB}}) \right)= \theta_A.
\end{equation}
{\it If} one could argue, that  as $\theta_A \to 0$ the sizes of
clusters will remain much smaller than the separation between them
$d \ll r_{AB}$, then we necessarily have $r_{AB} \to \infty$ and
Conjecture 1 follows. Unfortunately, in general the sizes of
clusters can grow as we change $\theta_a$'s, so this argument does
not always apply and one needs a more detailed knowledge of the
moduli space of solutions to (\ref{eq:integrability1}).

A related issue that we wish to address is a conjecture of
\cite{deBoer:2008fk}, relating multicentered solutions with single
$AdS_3 \times S^2$ near horizon geometry and attractor flow trees at
the  ``AdS point.'' The  ``AdS point'' is given by
\begin{equation}\label{eq:AdSpoint}
t_{AdS} = D^{AB} Q_B + i \infty P^A
\end{equation}
This is a point on the boundary of moduli space given by $\lim_{u
\to \infty}D^{AB} Q_B + i u P^A$ and we are considering limits of
attractor flows with $D^{AB} Q_B + i u P^A$ as an initial point.
Note that it is naturally selected by the near horizon limit
(\ref{eq:asympmoduli}). Note that the component of moduli space with
first split $D4 \to D4+D4$,  does not correspond to a single $AdS_3
\times S^2$, and this component also does not exist at the AdS
point. This motivated \cite{deBoer:2008fk} to suggest:

{\em \noindent  \textit{ {\bf Conjecture 2:} There is  a one to one
correspondence between $(i)$ components of the moduli space of
lifted multicentered solutions with a single AdS$_3 \times S^2$
asymptotic geometry and $(ii)$ attractor flow trees starting at the
AdS point.}}

We now give an argument in favor of this conjecture. As discussed in
Appendix A, the attractor tree is specified by the $H$-functions:

\begin{equation}
H(s^{(a)}) = \Gamma^{(a)} s^{(a)} - \Delta H^{(a)},
\end{equation}

\noindent where $s^{(a)}$ is the parameter along the flow on the
$a$-th edge. The rescaling  in (\ref{eq:family1}) leading to the
near horizon limit of \cite{deBoer:2008fk} results in changing the
$H$-functions to

\begin{equation}
H(s^{(a)}) \to H^{\lambda}(s^{(a)}) = \lambda^{3/2} \Gamma^{(a)} s^{(a)} - \Delta H^{(a)}_{\lambda}.
\end{equation}

\noindent According to (\ref{eq:DeltaHs}),  $\Delta H^{(a)}_{\lambda}$ depend linearly on and are completely determined in terms of $\Delta H_{\lambda}$, and $\Delta H_{\lambda} =  2 \Im (e^{-i \alpha} \Omega)_{t_{\infty}^{\lambda}}$, where $t_{\infty}^{\lambda} := B_{\infty}+ i \lambda J_{\infty}$. As the solution for the moduli (\ref{eq:AttrSol}) are homogeneous of degree zero in $H$, we can replace these $H^{\lambda}$-functions with:

\begin{eqnarray}\label{eq:defflow}
&& H^{\lambda}(s^{(a)}) \to \tilde H^{\lambda}(s^{(a)}) = \Gamma^{(a)} s^{(a)} - \Delta \tilde H_{(a)}^{\lambda}, \nonumber\\
&& \Delta \tilde H^{\lambda}  = \lambda^{-3/2} 2 Im(e^{-i \alpha} \Omega)|_{t_{\infty}^{\lambda}}.
\end{eqnarray}

We will refer to the split   flow defined by  (\ref{eq:defflow}) as
a $\lambda$-deformed flow. Note that for $\lambda$-deformed flows
the values of MS wall crossings parameters $s_{ms}^{(a) \, \lambda}$
in (\ref{s_ms}) will depend on $\lambda$. Our argument will be based
on two assumptions:

{\em \noindent  \textit{ {\bf Assumption 1:} There is a
$\lambda$-deformed version of the  SAFC. That is,  the components of
the  moduli space of $\lambda$-deformed solutions (\ref{eq:family1})
are in one to one correspondence with $\lambda$-deformed attractor
flow trees.}}

{\em \noindent  \textit{ {\bf Assumption 2:} The $\lambda$-deformed
solution ``survives" the near horizon limit, i.e. it corresponds to
an asymptotically AdS$_3 \times S^2$ geometry, iff the corresponding
$\lambda$-deformed attractor flow tree has all its flow parameters
$s_{ms}^{(a) \, \lambda}$ nonzero (and positive) in the limit
$\lambda \to \infty$. The attractor flow tree exists at the AdS
point iff all it's  flow parameters $s_{ms}^{(a)}$ stay nonzero (and
positive) as it's starting point approaches AdS point.}}

The second assumption is of course closely related to  Conjecture 1
above, because for the first split $D4 \to D4 + D4$ we have $s_{ms}
=   \frac{\sum_{a \in A} \theta_a}{\la \Gamma_A \Gamma_B \ra}$.
Given the above assumptions we want to prove that there is a one to
one correspondence between $\lambda$-deformed attractor flow trees,
that ``survive" the near horizon limit in the sense of Assumption 2,
and regular (not $\lambda$-deformed) attractor flow trees, that
start at the AdS point ( i.e. that have initial point approaching
this boundary point as $\lambda \to \infty$).

First, we note that the  first split of a $\lambda$-deformed flow
that ``survives'' the limit must be $D4 \to D6 + \overline{ D6}$. To
see this we use (\ref{FlowModuli}), to estimate the $\lambda$
dependence of $\Delta \tilde H^{\lambda}$:

\begin{equation}
\Delta \tilde H^{\lambda} = (\Delta \tilde H^0,\Delta \tilde H^A,\Delta \tilde H_A,\Delta \tilde H_0) \sim (\lambda^{-4},\lambda^{-2},\lambda^{-2},\lambda^{0}).
\end{equation}

\noindent From this we find that for $D4 \to D6 + \overline{ D6}$
the flow parameter of the first split is $s_{ms}^{\lambda} \sim
\lambda^0$, while for $D4 \to D4 +  D4$ it is $s_{ms}^{\lambda} \sim
\lambda^{-2}$. This means that only $D4 \to D6 + \overline{ D6}$ is
a valid split in the limit $\lambda \to \infty$.

For the chosen attractor trees we next look at the first edge of the
flow tree in the moduli space. Using formula A.6 from Appendix A,
the complexified K\"ahler moduli are:

\begin{eqnarray}\label{eq:moduliflow}
&& B^A_{\lambda}(s) =   D^{AB} \left(s P^C - \Delta \tilde H^C_{\lambda} \right) (s Q_B - \Delta \tilde H_B^{\lambda}) \nonumber\\
&& J^A_{\lambda}(s) =  (s P^A - \Delta \tilde H^A_{\lambda}) \sqrt{-6 ( s q_0 - \Delta \tilde H_0^{\lambda} - 1/2 Q^2(s) ) / (s P - \Delta \tilde H^0_{\lambda})^3}
\end{eqnarray}

\begin{figure}[htp]
\centering
\includegraphics[height=5cm,angle=0,trim=0 0 0 0]{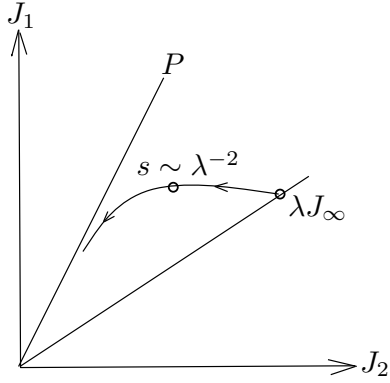}
\caption{The behavior of the flow for the first edge of the tree. }
\end{figure}

Figure 3 shows that the flow  starts at $t_{\infty}^{\lambda}$, but
for the flow parameter $s \sim \frac{1} {\lambda^2}$ the first term
in $(s P^A -  \Delta \tilde H^A_{\lambda})$ becomes comparable with
second term and then starts to dominate, so that the flow will go
along the $P$ direction. The transition from $J_{\infty}$
asymptotics to $P$ asymptotics occurs around $s \sim
\frac{1}{\lambda^2}$. Also note that the first split $D6\bar D6$
occurs long after this region at $s_{ms}^{\lambda} \sim \lambda^0$.

Now  choose a  value $\tilde s^{\lambda}$ of the flow parameter that
goes to zero more slowly than $ \frac{1}{\lambda^2}$, e.g. $\tilde
s^{\lambda} \sim \frac{1}{\lambda^{2-\epsilon}}$, with small
$\epsilon >0$.  From (\ref{eq:moduliflow}), it follows that
$J^A(\tilde s^{\lambda})$ will approach the $P$ direction as
$\lambda \to \infty$, and grow as $\lambda^{1-\epsilon/2}$, i.e.

\begin{equation}\label{eq:startpoint}
t^A(\tilde s^{\lambda}) \sim D^{AB}(P) Q_B (1+O(\lambda^{-\epsilon}))+i
 \lambda^{1-\epsilon/2} P^A \: const \: (1+ O(\frac{1}{\lambda})).
\end{equation}

We can think of the part of the attractor flow tree that starts at
$J^A(\tilde s^{\lambda})$ as a tree on its own. It is again
constructed in terms of $H$-functions, but now the $\Delta \tilde
H^{\lambda}$ function will look like:

\begin{equation}\label{eq:deltatildeH}
\Delta \tilde H^{\lambda} =\lambda^{-3/2} 2 Im(e^{-i \alpha}
\Omega)|_{t(\tilde s^{\lambda})}.
\end{equation}

The only difference of this  $\Delta \tilde H^{\lambda}$ with the
$\Delta H$ of the $\lambda$-undeformed flow with starting point
given by (\ref{eq:startpoint}), is the overall factor
$\lambda^{-3/2}$. Denoting the flow parameters for all edges of the
tree collectively by $s$, we can introduce new parameters $s' =
\lambda^{3/2} s$, in terms of which the $H$-functions will look like
the ones for the $\lambda$-undeformed flow with starting point given
by (\ref{eq:startpoint}). It follows from Appendix A that the
existence conditions, written in terms of parameters $s'$, are the
same as those written in terms of $s$, and furthermore the non-zero
$s_{ms}^{(a) \, \lambda}$ will correspond to non-zero
${s'}_{ms}^{(a) \, \lambda} $ since ${s'}^{(a) \, \lambda}_{ms} =
\lambda^{3/2} {s}^{(a) \,\lambda}_{ms}$. By virtue of Assumption 2,
the $\lambda$-deformed flow tree that "survives" the near horizon
limit has all its flow parameters $s_{ms}^{(a) \, \lambda}$
non-zero, and the corresponding $\lambda$-undeformed flow tree with
starting point (\ref{eq:startpoint}) exists at the AdS point.

In order to prove Conjecture 2 in the other  direction consider a
family of attractor flow trees whose initial point approaches the
AdS point. Note that only the trees with the first split $D4 \to D6
\overline{ D6}$ exist in this limit, as shown in
\cite{deBoer:2008fk}, eq.(3.64). Without loss of generality, for
sufficiently large $\lambda$  we can choose the initial points to be
given by the right-hand side of (\ref{eq:startpoint})  for some
$t_{\infty}$.  Now, due to Assumption 2, the existence of the
attractor flow tree at the AdS point means that in the limit
$\lambda \to \infty$ all the flow parameters of these trees,
${s'}_{ms}^{(a) \, \lambda}$, stay non-zero. The dependence on
$\lambda$ in ${s'}_{ms}^{(a) \, \lambda}$ originates from the
dependence in the starting point (\ref{eq:startpoint}). We can use
the discussion above to argue that there exists a corresponding
$\lambda$-deformed flow tree, starting at $t_{\infty}$ and passing
through the point (\ref{eq:startpoint}) at some parameter $\tilde
s^\lambda$. For this $\lambda$-deformed flow tree to "survive" the
limit $\lambda \to \infty$ we must have all ${s}_{ms}^{(a) \,
\infty}$ non-zero and positive, due to Assumption 2. As the relation
between the flow parameters for the two trees is $s^{(a) \,
\lambda}_{ms} = \lambda^{-3/2} {s'}^{(a) \, \lambda}_{ms}$, some of
the $s^{(a) \, \lambda}_{ms}$ of the $\lambda$-deformed flow might
go to zero in the limit $\lambda \to \infty$, leading to trouble. We
will now argue that in fact this cannot  happen. To this end, first
introduce a notation, analogous to the one in
(\ref{eq:integrability1}):
\begin{equation}
\theta(\Gamma):= 2 \Im (e^{-i \alpha} Z(\Gamma))_{\infty}
\end{equation}
 According to (\ref{s_ms}), for
 each edge $a$ the flow parameter $s_{ms}^{(a)}$ is given
 by a linear combination, with rational coefficients,
  of $\theta(\Gamma_i)$, where $i$ runs over all
  the intermediate charges occuring in the   path from the root of the tree
  to the edge $a$. For the $\lambda$-deformed flow
   these $\theta(\Gamma_i)$ have a definite scaling under $\lambda$-scaling.
   For instance, since the first split is always $\Gamma(D4) \to \Gamma_1+\Gamma_2$
   where $\Gamma$ and $\Gamma_2$ have nonzero (and opposite) $D6$ charge,
   we have $\theta(\Gamma_1) = -\theta(\Gamma_2) \sim \lambda^0$ and
$\theta(\Gamma_1)$ will enter the expressions for all ${s'}^{(a)
\lambda}_{ms}$. Other $\theta(\Gamma_i)$ will in general have
$\CO(\lambda^0)$ scaling (i.e. those with nonzero $D6$ charge) but,
examining examples, we find that the coefficient of the $\lambda^0$
term will be some complicated nonlinear expression in terms of the
intersection products of the charges, which does not vanish in these
examples and hence we expect does not vanish generically. For
example for figure 4, $s_{ms}^{(4)}$ for the edge with $\Gamma_4$,
it is a combination of the form:
\begin{equation}
s_{ms}^{(4)} = \frac{ \theta_5
- \frac{\langle \Gamma_5, \Gamma_2 \rangle}{\langle \Gamma_3, \Gamma_4 \rangle} \theta_3 + \frac{\langle \Gamma_3, \Gamma \rangle \langle \Gamma_5, \Gamma_2 \rangle}{\langle \Gamma_1, \Gamma_2 \rangle \langle \Gamma_3, \Gamma_4 \rangle} \theta_1 - \frac{\langle \Gamma_5, \Gamma \rangle}{\langle \Gamma_1, \Gamma_2 \rangle} \theta_1}{\langle \Gamma_5, \Gamma_6 \rangle}
\end{equation}
Here $\theta_5 \sim \lambda^{-2}$, $\theta_1 \sim \lambda^{0}$,
$\theta_3 \sim \lambda^{0}$. If we assume that all $D6$ branes have
$D6$ charges $\pm 1$, then in the limit $\lambda \to \infty$
$\theta_1=-\theta_3$,  the leading coefficient of  $s_{ms}^{(4)}$ is proportional to
\begin{equation}
-\langle \Gamma_3, \Gamma_6 \rangle \langle \Gamma_5, \Gamma_1
\rangle + \langle \Gamma_1, \Gamma_5 \rangle \langle \Gamma_5,
\Gamma_6 \rangle + \langle \Gamma_1, \Gamma_6 \rangle \langle
\Gamma_5, \Gamma_6 \rangle + \langle \Gamma_1, \Gamma_6 \rangle
\langle \Gamma_5, \Gamma_3 \rangle
\end{equation}
which has no reason to vanish. In this way we can argue  that all
${s'}^{(a) \, \lambda}_{ms}$ will have an order $\sim \lambda^0$
contribution whose coefficient  will not scale to zero as $\lambda
\to \infty$, at least not in general.

\begin{figure}[htp]
\centering
\includegraphics[height=5cm,angle=0,trim=0 0 0 0]{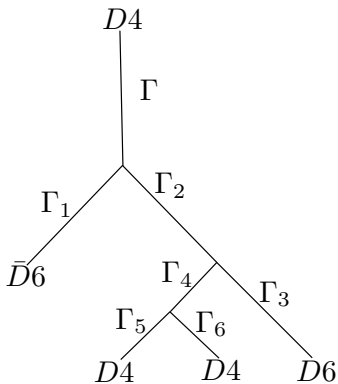}
\caption{An example of attractor flow tree. }
\end{figure}\label{fig:exampletree}

To summarize, we have shown that there  is a one to one
correspondence between $\lambda$-deformed attractor flow trees that
``survive" the near horizon limit, and regular attractor flow trees,
starting at AdS point. If one grants Assumptions 1 and 2 this would
actually  prove   Conjecture 2, and hence the conjecture of
\cite{deBoer:2008fk}.

\acknowledgments

We would like to thank F. Denef, E. Diaconescu,   D. Jafferis, and
D. Van den Bleeken for useful discussions and remarks. We are
grateful to F. Denef for useful comments on a preliminary draft of
this paper. This work is supported by the DOE under grant
DE-FG02-96ER40949.

\appendix

\section{Attractor flow trees}\label{AttrFlTr}

In this appendix we summarize some facts about attractor flow trees.

Consider type IIA string theory on $M_4 \times X$, where $X$ is a
Calabi-Yau of generic holonomy. BPS states in the theory are labeled
by their electromagnetic charges $\Gamma = (p^0, p^a, q_a, q_0)$.

The low energy theory is $N=2$ supergravity coupled to $n_V =
h^{1,1}(X)+1$ vector multiplets representing complexified K\"ahler
moduli of X. In this low energy theory BPS states are realized as
single or multicentered black hole solutions.

It was conjectured in \cite{Denef:2000nb,Denef:2000ar}   that the
existence of multicentered BPS solutions of supergravity can be
analyzed in terms of the the existence of split attractor flow
trees. Some attempts at making this conjecture more precise were
made in \cite{Denef:2007vg,deBoer:2008fk}.

\bigskip
\textbf{Split Attractor Flow Conjecture (SAFC)}:

\begin{enumerate}

\item[a)] The components of the moduli spaces (in $\vec x_i$) of the
multicentered BPS solutions with constituent charges $\Gamma_i$ and
background $t_\infty$,
 are in 1-1 correspondence
with the attractor flow trees beginning at $t_\infty$ and
terminating on
  attractor points for $\Gamma_i$.

\item[b)] For a fixed $t_\infty$ and total charge $\Gamma$ there are only
a finite number of attractor flow trees.

\end{enumerate}

A practical recipe of identifying the intervals with the corresponding tree topologies is the following:   tune the moduli at infinity such that they  approach the first
MS wall of a given attractor flow tree. Then, as we change the moduli across that MS wall, the corresponding component of moduli space of solutions to  (\ref{eq:integrability1}) ceases to exist. In this paper we assume
the truth of the split attractor flow conjecture and simply
establish the existence of attractor flow trees.

 We now give an explicit description of an attractor flow tree.

First, we introduce some notation. For a general tree we denote
quantities, related to particular vertex, by $X^{(\vec \epsilon)}$
for quantity $X$. Here $\vec \epsilon$ is a vector of $+$ and $-$
signs and the sequence of $+$ and $-$ corresponds to sequence of
right and left turns that one needs to make when going from the
origin of the tree to that vertex (the origin itself will have no
superscript).

The attractor equation for the edge starting at vertex $(a)$, looks
like:

\begin{equation}\label{attr_eq}
    2 e^{-U} Im(e^{-i \alpha^{(a)}} \Omega(t)) = -H(s^{(a)}),
\end{equation}

\noindent where $\Omega(t) = -\frac{1}{\sqrt{4/3 J^3}}e^{B+iJ}$ (in IIA picture), $e^{U}$ is the metric warp factor, $\alpha^{(a)}$ is the phase of central charge $Z(\Gamma^{(a)})$,
$s^{(a)}$ is a parameter of the flow on this edge, and
\begin{equation}
H(s^{(a)}) = \Gamma^{(a)} s^{(a)} - \Delta H^{(a)}.
\end{equation}

$\Delta H^{(a)}$ depends only on the moduli   at infinity and is
determined recursively by summing contributions from the origin of
the tree up to vertex $(a)$:

\begin{eqnarray}\label{eq:DeltaHs}
&& \Delta H = 2 Im(e^{-i \alpha} \Omega)|_{t_{\infty}}\nonumber\\
&&  \Delta H^{(+)} = \Delta H^{(-)}= \Delta H  - \Gamma s_{ms} \nonumber\\
&&  \Delta H^{(++)} = \Delta H^{(+-)} =\Delta H^{(+)} - \Gamma^{(+)} s^{(+)}_{ms} \nonumber\\
&&  \Delta H^{(-+)} = \Delta H^{(--)} =\Delta H^{(-)} - \Gamma^{(-)}
s^{(-)}_{ms} ...
\end{eqnarray}

\noindent where $s^{(a)}_{ms}$ are values of parameters along the
flow, for which surfaces of marginal stability are crossed:

\begin{eqnarray}\label{s_ms}
&& s_{ms} = \frac{\la \Gamma^{(+)} \Delta H \ra}{\la \Gamma^{(+)} \Gamma \ra}\nonumber\\
&& s^{(+)}_{ms} = \frac{\la \Gamma^{(++)} \Delta H^{(+)}\ra }{\la \Gamma^{(++)} \Gamma^{(+)} \ra}\nonumber\\
&& s^{(-)}_{ms} = \frac{\la \Gamma^{(-+)} \Delta
H^{(-)}\ra}{\la \Gamma^{(-+)} \Gamma^{(-)}\ra}...
\end{eqnarray}

The solution to the attractor equations (\ref{attr_eq}), that is,
the image of the flow in moduli space, can be written in closed form
in terms of the entropy function $S(p,q)$ \cite{Denef:2003}:

\begin{equation}\label{ModuliSolGen}
 t^A(s^{(a)}) = \left.
 \frac{\frac{\partial S}{\partial q_A} + \pi i p^A}{\frac{\partial S}{\partial q_0}-\pi i
 p^0} \right|_{(p,q)=H( s^{(a)})}.
\end{equation}

Here, the parameter $s^{(a)}$ varies as: $s^{(a)} \in (0, \infty)$
for the terminal edge, and $s^{(a)} \in (0,s^{(a)}_{ms} )$ for an
inner edge.

For a given attractor tree to exist, all its edges have to exist.
Terminal edges exist if the discriminants of terminal charges are
positive,   or if the terminal charge is pure electric or magnetic,
which corresponds to the flow going to the boundary of moduli space.
Inner edges exist if:
\begin{enumerate}
\item The flow reaches the MS wall at a positive flow parameter $s^{(a)}_{ms} > 0 $
\item And, an MS wall (not an anti-MS wall) is crossed, i.e.
$\frac{Z(\Gamma^{(a+)})}{Z(\Gamma^{(a-)})}|_{s^{(a)}_{ms}} > 0$
\item And, the MS wall is crossed before the flow hits a zero of the central charge (if
present):
$$s^{(a)}_{ms} \le s^{(a)}_0 \quad {\rm or} \quad s^{(a)}_0 \le 0$$
\noindent where $s^{(a)}_0$ is the value where the flow crashes on a
zero.
\end{enumerate}

\noindent For a D4-D2-D0 charge we give explicit formulae for
attractor flow in moduli space:

\begin{eqnarray}\label{FlowModuli}
&& t^a(s) = D(P(s))^{ab} Q_b(s) + i P^a(s) \sqrt{-6 \hat q_0(s) / P^3(s)}\nonumber\\
&& \Gamma(s) = p^0(s)+P(s)+Q(s)+q_0(s) dV = s \Gamma - \Delta H \nonumber\\
&&\Delta H = \frac{2 Im(\bar Z \Omega)}{|Z|}|_{\infty}  = \nonumber\\
&& = \frac{2}{\sqrt{\frac{4}{3} J^3} }
 \left(2 \frac{-Q \cdot J + P \cdot B \cdot J}{P \cdot J^2}  -J + J^2 \frac{-Q \cdot J + P \cdot B \cdot J}{P \cdot J^2} + \frac{J^3}{6}\right)|_{\infty} \nonumber\\
\end{eqnarray}

In the formula for $\Delta H$ we used the large $J_{\infty}$ approximation and dropped relative corrections of order $O(J^{-2}_{\infty})$. The expression for $t^a(s)$ was found  from (\ref{ModuliSolGen})
putting $p^{0}(s)=0$. Strictly speaking, this is not true because
already $\Delta H$ contains non-zero contribution to $p^{0}(s)$. To
estimate the error that we make, take the expression for the moduli
for a 1-parameter moduli space and expand it around $p^{0}(s)=0$.
The first correction looks like:

\begin{equation}\label{ModCorr}
\delta_1 t(s) =\left[  \frac{2 Q(s)^2 - 3 P(s) q_0(s)}{P(s)^3}   + i
\frac{\sqrt{3} P(s) Q(s) (2 Q(s)^2 -3  P(s) q_0(s))}{3 P(s)^3
\sqrt{P(s)^2 (Q(s)^2 - 2 P(s) q_0(s))}}\right] p^0(s)
\end{equation}

Focusing on $J_{\infty}$ dependence, $\Gamma(s)$  in (\ref{FlowModuli}) can be written as

\begin{equation}
\Gamma(s) = \left( O(J_{\infty}^{-5/2}) , s P  + O(J_{\infty}^{-1/2}) , s Q + O(J_{\infty}^{-1/2}) , s q_0+ O(J_{\infty}^{3/2}) \right)
\end{equation}

This means that, for instance, for $s$ of order 
$s  \sim J_{\infty}^{-1/2+\epsilon}$ with $0 \le \epsilon \le 2$
(which covers all the cases of interest in this paper) the correction in (\ref{ModCorr}) is of order

\begin{equation}
\delta_1 t(s) \sim O(J_{\infty}^{-2 \epsilon}) + i  O(J_{\infty}^{-1 -3/2 \epsilon})
\end{equation}

\noindent and can be neglected in large $J_{\infty}$ limit.

\end{document}